\documentclass[11pt,twoside,a4paper]{article}
\usepackage[utf8]{inputenc}
\usepackage{graphicx,color}
\usepackage{times}
\usepackage{xspace}
\usepackage{booktabs}
\usepackage[english]{babel}
\usepackage[bf,font=small]{caption} 
\usepackage{amssymb}
\usepackage{cite}
\usepackage{xspace}
\usepackage[top=1.135in,bottom=1.17in,left=1.05in,right=1.05in]{geometry}
\usepackage{amsmath,multicol}
\usepackage{subcaption}

\usepackage[colorlinks=true, pdftex, citecolor=blue, urlcolor=blue, linkcolor=blue]{hyperref}
\usepackage{lineno}

\newcommand{\mrm}[1]{\ensuremath{\mathrm{#1}}\xspace}

\newcommand{\reject}{\ensuremath{\mathrm{rej}}\xspace}
\newcommand{\cns}{\ensuremath{c_{\mbox{\tiny NS}}}\xspace}

\newcommand{\MeV}{\ifmmode {\mathrm{\ Me\kern -0.1em V}}\else
                   \textrm{Me\kern -0.1em V}\fi}%
\newcommand{\GeV}{\ifmmode {\mathrm{\ Ge\kern -0.1em V}}\else
                   \textrm{Ge\kern -0.1em V}\fi}%
\newcommand{\TeV}{\ifmmode {\mathrm{\ Te\kern -0.1em V}}\else
                   \textrm{Te\kern -0.1em V}\fi}%

\newcommand{\eqRef}[1]{eq.~(\ref{#1})\xspace}

\newcommand{\eqsRef}[1]{eqs.~(\ref{#1})\xspace}

\newcommand{\secRef}[1]{section~\ref{#1}\xspace}

\newcommand{\secsRef}[1]{sections~\ref{#1}\xspace}

\newcommand{\figRef}[1]{fig.~\ref{#1}\xspace}

\newcommand{\inst}[1]{$^{#1}$}
\renewcommand{\and}{, }

\title{Automated Parton-Shower Variations in Pythia 8}
\author{S. Mrenna\inst{1} and P. Skands\inst{2}\\[5mm]
\normalsize\parbox[c]{0.85\textwidth}{
$^1$: Computing Division, Fermilab, Batavia, Illinois, USA \\
$^2$: School of Physics and Astronomy, Monash University, VIC-3800, Australia\\
}}
\date{}

\begin{document}
\maketitle
{\abstract{
In the era of precision physics measurements at the LHC, efficient and exhaustive estimations of theoretical uncertainties play an increasingly crucial role. In the context of Monte Carlo (MC) event generators, the estimation of such uncertainties traditionally requires independent MC runs for each variation, for a linear increase in total run time. In this work, we report on an automated evaluation of the dominant (renormalization-scale and non-singular) perturbative uncertainties in the PYTHIA~8 event generator, with only a modest computational overhead. Each generated event is accompanied by a vector of alternative weights (one for each uncertainty variation), with each set separately preserving the total cross section. Explicit scale-compensating terms can be included, reflecting known coefficients of higher-order splitting terms and reducing the effect of the variations. The formalism also allows for the enhancement of rare partonic splittings, such as $g\to b\bar{b}$ and $q\to q\gamma$, to obtain weighted samples enriched in these splittings while preserving the correct physical Sudakov factors. 
}}

\tableofcontents

\section{Introduction}

Event generators~\cite{Buckley:2011ms} are used in almost all tests of the 
Standard Model at colliders.  The current state-of-the-art allows for
fixed-order corrections to the matrix elements used in these
predictions and a consistent matching or merging between the matrix
element and parton shower contributions.
The components of such predictions are based on approximation, and it
is necessary to estimate their reliability.   For the fixed--order
(matrix element) piece of these calculations,  estimates of the
uncertainty come from varying a common factorization ($\mu_F$) and
renormalization ($\mu_R$) scale and simultaneously considering the
correlation with the parton distribution functions (PDFs).  While
these uncertainty estimates may be tedious to compute, it is a
relatively simple procedure to calculate the relative weight of 
an event after scale and PDF variations.   There is an additional
simplicity to these uncertainty estimates.   The rest of the event
development is usually unchanged by these variations.  Therefore,
each particle-level prediction can be recycled after applying the
weight correction for that event.    This is particularly important
if the entire event has been folded with a (time-consuming) detector
simulation.

The uncertainty on the other components of the prediction, such as
parton showering, multiparton interactions, and hadronization, is
more challenging to estimate.   This is because of the algorithms
applied to sample probability distributions and the iterative nature
of the algorithms.   The state-of-the-art is to select a
(small) number of event-generator parameters and make entirely new
predictions based on them, as e.g.\ in the ``Perugia'' tune variations~\cite{Skands:2010ak} and/or ``eigentune'' variations~\cite{Buckley:2009bj,Richardson:2012bn,ATL-PHYS-PUB-2014-021,Khachatryan:2015pea}.  Since each of these new predictions
makes different particle-level predictions, each generated event must
be passed through a detector simulation as part of a realistic analysis.  
This fact greatly reduces the number of parameter variations than can be performed.

In this paper, we present a method to estimate the effect of
parameter variations in the parton shower for a given kinematic
configuration.    This is similar to what was done previously
in VINCIA for final-state radiation (FSR)~\cite{Giele:2011cb}, 
but is here extended to initial state radiation (ISR) and adapted to PYTHIA's parton-shower framework~\cite{Sjostrand:2004ef,Corke:2010yf,Sjostrand:2014zea}. 

We also show how to use the same method to
generate a weighted sample enhanced in the occurrence of specific 
shower branchings, such as $g\to b\bar{b}$, with correctly calculated weights (including
correct physical Sudakov form factors). This could be useful, e.g., for the B physics community. (We note that equivalent proposals for ``biasing'' or ``boosting'' specific shower splitting probabilities were also made in \cite{Hoeche:2009xc,Lonnblad:2012hz}.)
The two methods can be combined, so one can also
get uncertainties on a biased sample, although this latter capability has not yet been implemented in the current PYTHIA code. 

\section{Method}

The probability for a branching in the parton shower is encapsulated
in the Sudakov form factor which, for realistic applications, must be evaluated numerically.   A practical numerical method for 
this is the veto algorithm, known in computer science as the
"thinning algorithm"~\cite{Lewis:1979,devroye2013non}, which
involves rejecting (or thinning out) trial branchings. We start with a brief review of this method as applied to parton showers in \secRef{sec:proof}. We then turn to the main focus of this paper: incorporating systematic variations of the branching probabilities. For a unitary (probability-conserving) shower, such variations necessarily imply opposite variations in the non-branching  probabilities through the rejections. The specific form these variations must have to preserve the unitarity of the shower are derived in \secRef{sec:variations}. A further interesting application of the same framework is presented in \secRef{sec:bias}, allowing to generate correctly weighted showers with biased kernels, as was already proposed for $q\to q\gamma$ splittings in~\cite{Hoeche:2009xc}. With the general formalism now in hand, \secsRef{sec:muRfac} and \ref{sec:cNS} describe the specific application of the framework to renormalisation-scale and non-singular term variations in the shower, respectively. 

At the technical level, in the original VINCIA implementation~\cite{Giele:2011cb}, the set of variations that could be performed were defined by the authors (hardcoded), with limited options for users to modify e.g.\ by which factor to vary the renormalisation scale up and down. The PYTHIA implementation has been made significantly more general,  allowing users considerable flexibility to define any number of simultaneous or separate variations, as documented in detail in PYTHIA's online HTML documentation, with a set of default variations chosen by the authors. 

As in VINCIA, the modification to the rejection and acceptance probabilities are accumulated during the shower evolution and presented after the shower has finished as (a set of) alternative global event weights; one for each variation. The relative probability for each event to occur under different showering assumptions (represented by the variations) is given by the weight calculated for the given variation relative to the nominal (unvaried) event weight. 

We note that, since unitarity is strictly imposed on these variations, each set of weights should integrate to the same total cross section. Bear in mind, however, that this will only really be true in the limit of infinitely many events. Depending on the magnitude of each variation and how ``long'' the shower evolutions are (bigger phase spaces imply more room for changes to accumulate), the variation weights will fluctuate around their mean values. This will reduce the statistical precision on the uncertainty variations relative to the nominal sample. To exemplify, take a sample of 100 identical hard 4-jet events, and say that one of them experienced a very unlikely branching somewhere deep in the shower, say at the 20th branching (i.e.\ with minimal impact on 4-jet distributions). These 100 events would all enter with the same weight in the nominal sample. But the event that happened to contain the unlikely 20th branching can acquire a much larger weight in one of the variations if the probability for that branching to occur is much larger for that variation. The 4-jet cross section computed from the variation weights would then be dominated by the single event with large weight, corresponding to a much worse statistical precision, in spite of the fact that the actual weight change occurred not at the 4-jet level but much deeper in the shower. This is a simple  consequence of accumulating the variations through the shower history, which --- depending on future uses of the algorithm --- may make it desirable to introduce further options for controlling the amount of variation performed at each stage of the shower. For the time being, for practical applications, we advise to monitor the variations of the uncertainty weights in each histogram bin so that any issues due to very rare events with very large variation weights do not go unnoticed. 

\subsection{Proof of the Standard Veto Algorithm \label{sec:proof}}

Given a differential branching probability, $P(t,z)$,
with $t\propto Q^2$ the shower evolution variable and $z$ a complementary phase-space invariant (which in the DGLAP picture can be identified with the collinear energy-sharing fraction), 
a standard parton-shower algorithm generates the scale of the next branching by solving the following equation for $t$,
\begin{equation}
{\cal R}_t ~=~\Delta(t_0,t) ~=~\exp\left(-\int_t^{t_0} dt_1 \int dz_1 P(t_1,z_1)\right)~
,\label{eq:Sudakov}
\end{equation}
with $t_0$ the starting scale for the evolution, ${\cal R}_t\in[0,1]$ a uniformly distributed random number, and  $\Delta$ the Sudakov factor, or no-branching probability. 
In the specific case of PYTHIA's transverse-momentum-ordered showers~\cite{Sjostrand:2004ef}, the differential branching probability is 
\begin{equation}
P(t,z) = \frac{\alpha_s(t)}{2\pi}\frac{P(z)}{t}~,
\end{equation}
with $t=p_{\perp\mrm{evol}}^2$~\cite{Sjostrand:2004ef}, and $P(z)$ a DGLAP splitting kernel~\cite{Gribov:1972ri,Dokshitzer:1977sg,Altarelli:1977zs}. 
We emphasize, however, that the formalism presented here is valid for arbitrary $P(t,z)$ and could be applied equally well to dipole/antenna-showers. 

After the selection of $t$, a value for $z$ is then selected according to a second random number, by solving for $z$ in the following equation
\begin{equation}
{\cal R}_z ~=~ \frac{\int_{z_\mrm{min}}^z dz_1 \ P(t,z_1)}
{\int_{z_\mrm{min}}^{z_\mrm{max}} dz_1 P(t,z_1)}~, \label{eq:Rz}
\end{equation}
with ${\cal R}_z\in[0,1]$ a different uniformly distributed random number.
This generates the resummed probability distribution,
\begin{equation}
\frac{d{\cal P}}{dt\ dz} = P(t,z) \ \Delta(t_0,t)~.\label{eq:resummed}
\end{equation}
However, since $P(t,z)$ can be complicated to integrate\ (especially in the presence of matrix-element or higher-order corrections to $P$) and \eqsRef{eq:Sudakov} \& \eqref{eq:Rz} can be difficult to invert analytically for $t$ and $z$, a simple and powerful trick is normally used to transform the problem: the ``veto algorithm''. Instead of using the exact $P$ in \eqsRef{eq:Sudakov} \& \eqref{eq:Rz}, one instead uses a simpler ``trial'' overestimate, $\hat{P}(t,z) > P(t,z)$, constructed specifically such that it can be easily integrated and inverted. (The integration boundaries in $z$ can also be extended to cover a larger region than the physical one, though such details are not important here.) 
Trial branchings generated according to $\hat{P}$ (i.e. with a Sudakov $\hat{\Delta}$ based pm $\hat{P}$) are then accepted with the probability
\begin{equation}
P_\mrm{acc}(t,z) = \frac{P(t,z)}{\hat{P}(t,z)}~,
\label{eq:Pacc}
\end{equation}
with $P(t,z)=0$ outside the boundaries of the physical phase space, and $P_\mrm{acc}<1$ 
guaranteed by $\hat{P}>P$. 
If the trial is accepted, physical momenta are generated corresponding to the chosen values of 
$t$ and $z$, and the pre-branching partons are replaced by the post-branching ones, including
the effects of recoils etc. If the trial is rejected (with probability $P_\reject = 
1-P_\mrm{acc}$), the parton system remains in its original state. In either case, the scale of 
the (accepted or rejected) trial becomes the new value for $t_0$, from which the evolution is 
restarted to find the next (lower) trial scale.  The procedure ends when $t<t_{min}$.

Before considering how to modify this algorithm to produce
uncertainty variations and bias weights, we first demonstrate the
all-orders proof of why the veto algorithm does end up producing the
correct form of the physical resummed distribution,
\eqRef{eq:resummed}.  This will be useful as the main starting point
below, and is often neglected in the literature.\footnote{The oldest
equivalent explicit treatment we are aware of in the particle-physics
literature is given by Sj\"ostrand and van Zijl in the context of
their Sudakov-based approach to multiple-parton
interactions~\cite{Sjostrand:1987su},  proving that the sum over
$p_\perp$-ordered MPI reproduces the naive (inclusive, unordered)
cross sections when summed over all possible orderings. The standard
veto algorithm applied to parton showers is also described in
Ref.~\cite{Sjostrand:2006za} and is further discussed in
Ref.~\cite{Platzer:2011dq,Platzer:2011dr,Lonnblad:2012hz}.   However, as mentioned above, the
basics of the algorithm can be found in references
~\cite{Lewis:1979,devroye2013non}.}

Consider the probability distribution for the first accepted branching. For the specific case 
of zero rejected trials preceding it (suppressing the $z$ dependence for clarity), it is:
\begin{equation}
\frac{d{\cal P}_0}{dt} = P_\mrm{acc}(t) \ \hat{P}(t) \  \hat{\Delta}(t_0,t) = P(t)\ 
\hat{\Delta}(t_0,t)~,
\end{equation}
hence for zero rejected trials the accept-probability factor results in the correct $P(t)$ but
it is associated with the wrong (trial) Sudakov factor, $\hat{\Delta} < \Delta$. For one 
rejected trial preceding the accepted branching, we have:
\begin{eqnarray}
\frac{d{\cal P}_1}{dt} &=& P(t) \ \int_{t}^{t_0} dt_1  
\hat{P}(t_1)\left(1-P_\mrm{acc}(t_1)\right) \hat{\Delta}(t_0,t_1) \hat{\Delta}(t_1,t) \\
 &=& P(t) \ \hat{\Delta}(t_0,t) \int_{t}^{t_0} dt_1  \left(\hat{P}(t_1)-P(t_1)\right) ~,
\end{eqnarray}
where we have used the fact that Sudakov products combine trivially when the underlying system 
doesn't change, $\Delta(t_1,t_2) = \Delta(t_1,t)\Delta(t,t_2)$ and we again find the ``wrong'' 
(trial) Sudakov but now it is accompanied by a factor that depends explicitly on the difference
between $\hat{P}$ and $P$. For two rejected trials preceding the accepted branching, 
\begin{equation}
\frac{d{\cal P}_2}{dt} 
 = P(t) \ \hat{\Delta}(t_0,t) \int_{t}^{t_0} dt_1  \left(\hat{P}(t_1)-P(t_1)\right) 
 \int_{t}^{t_1} dt_2 
 \left(\hat{P}(t_2)-P(t_2)\right)~,\label{eq:P2}
\end{equation}
and similarly for $n$ rejected trials preceding the accepted branching. The crucial point
in the proof is 
to recognize that the double integral in \eqRef{eq:P2} is in fact a triangle integral with a 
factorized integrand symmetric under interchange of the two integration variables, hence it can
be written
\begin{equation}
\int_0^1 dx f(x) \int_0^x dy f(y) = \frac12 \int_0^1 dx f(x) \int_0^1 dy f(y) = \frac12 \left(\int_0^1 dx f(x)\right)^2~,
\end{equation}
and similarly for the higher-$n$ terms which yield hyper-triangle integrals that can always be written on product form, prefaced by a factor $1/n!$ which gives the fractional volume occupied by a single ordered slice $t_0 > t_1 > t_2 > \ldots > t_n > t$ of the full $n$-hypercube.

Thus, the densities for each possible number of rejected trial branchings form nothing but the terms of an expanded exponential. The sum over all possible numbers of preceding failed trial branchings is therefore,
\begin{align}
\frac{d{\cal P}}{dt} &= P(t)\ \hat{\Delta}(t_0,t) \left[
1 + \sum_{n=1}^{\infty} \frac{1}{n!}\left(\int_t^{t_0} dt_1 \left(\hat{P}(t_1) - P(t_1)\right)\right)^n\right] \\
&= P(t)\ \exp\left(-\int_t^{t_0} dt_1\hat{P}(t_1)\right) \exp\left(\int_t^{t_0} dt_1\hat{P}(t_1) - P(t_1)\right) \\
&= P(t)\ \exp\left(-\int_t^{t_0} dt_1{P}(t_1)\right)~,\\
&= P(t)\ \Delta(t_0,t)~,
\end{align}
where we inserted the definition of the trial Sudakov, $\hat{\Delta}$, in the second line, cancelled it against the $\hat{P}$ term from the failed-branching exponential, and finally used the definition of the physical Sudakov, \eqRef{eq:Sudakov}. The last line is the desired expression, which now gives the physical resummed branching probability, independently of the trial function. This expression is identical to \eqRef{eq:resummed}, proving the correctness of the veto algorithm and in particular that the final result is independent of the choice of trial function, as long as $\hat{P} > P$. 

\subsection{Veto Algorithm with Uncertainty Variations \label{sec:variations}}

The main part of our paper consists of the proof, to all orders in perturbation theory, of a conjecture developed by one of us in Ref.~\cite{Giele:2011cb} in the context of the VINCIA shower generator~\cite{Giele:2007di}. According to this proposal, the veto algorithm discussed above can be modified to simultaneously compute several alternative sets of weights for each event, answering roughly: what \emph{would} the weight of this event have been, \emph{if} we had used, for instance, an alternative value for the strong coupling or an alternative splitting function? The number of variations that can be included is in principle infinite (each requiring very little computing and memory resources), hence several alternative  definitions of the same source of uncertainty can be evaluated simultaneously (e.g., renormalisation-scale variations by factors $\sqrt{2}$, $2$, and $4$ can all be included) and final plots can be made using only a subset of these. We here prove the validity of the algorithm to all orders in perturbation theory, and implement it in the PYTHIA~8 event generator~\cite{Sjostrand:2014zea}.

Consider a parton shower based on the veto algorithm discussed above, with the physical trial-accept probability, $P_\mrm{acc}$, given by \eqRef{eq:Pacc}. Consider further an alternative shower algorithm, defined by a different physical trial-accept probability, $P'_\mrm{acc}$, 
\begin{equation}
P_\mrm{acc}'(t,z) = \frac{P'(t,z)}{\hat{P}(t,z)}~,
\end{equation}
where the difference between the alternative radiation kernel $P'$ and the original $P$ can be, for instance, different $\alpha_s$ scale choices, different non-singular terms in the splitting kernels, and/or different effective higher-order contributions to the splitting kernels. Note however that we assume that the $t$ and $z$ definitions remain the same. Translations between different $t$ choices are discussed in~\cite{Hartgring:2013jma} (and the resulting equations are used in VINCIA to provide an uncertainty variation corresponding to the difference between virtuality-ordered and $p_\perp$-ordered showers) while exploring different $z$ definitions (and more generally, different recoil strategies) would require a future generalisation of the algorithm presented here.

The proposal to compute the probability of an event generated by $P'$ based on an event 
generated using $P$ is as following~\cite{Giele:2011cb} (suppressing again the $z$ dependence 
for clarity):
\begin{enumerate}
\item Start the event evolution by setting all weights (nominal and uncertainty-variation ones) equal to the input weight of the event, $w'=w$.
\item If the trial branching is accepted, multiply the alternative weight $w'$ by the relative ratio of accept probabilities,  
\begin{equation}
 R_\mrm{acc}'(t)~=~\frac{P'_\mrm{acc}(t)}{P_\mrm{acc}(t)}~=~ \frac{P'(t)}{P(t)}~.
\end{equation}
\item If the trial branching is rejected, multiply the alternative weight $w'$ by the relative ratio of reject-probabilities, 
\begin{equation}
R_\reject'(t)~=~\frac{P'_\reject(t)}{P_\reject(t)}~=~
 \frac{1-P'_\mrm{acc}(t)}{1-P_\mrm{acc}(t)}
~=~\frac{\hat{P}(t)-P'(t)}{\hat{P}(t)-P(t)}~.
\label{eq:Rreject}
\end{equation}
\item If desired, the detailed balance between the accept and reject probabilities could optionally be allowed to be broken by up to a non-singular term, $P'_\mrm{acc} \neq 1-P'_\reject$, to represent uncertainties due to genuine (non-canceling) higher-order corrections, which would modify the total cross sections. For the current implementation in PYTHIA, however, we do not consider this possibility further. 
\end{enumerate}
Step 2 is responsible for adjusting the naive splitting probabilities, while Step 3 is responsible for adjusting  the no-splitting Sudakov factors. The result is that the set of weights $w'$ represents a separately unitary event sample, with $\left<w'\right> = \left<w\right>$; i.e., the samples integrate to the same total cross section. We already know that the probability distribution of the generated event sample, when applying the nominal set of weights, $w$, is the distribution defined by \eqRef{eq:resummed}. We shall now prove that the probability distribution obtained from the same generated event sample, when applying the set of weights $w'$, is the correct resummed distribution for the $P'$ radiation kernels, 
\begin{equation}
\frac{d{\cal P}'}{dt\ dz} = P'(t,z) \ \Delta'(t_0,t)~,\label{eq:resummedPrime}
\end{equation}
where the apostrophes on both $P'$ and $\Delta'$ emphasize that the modified radiation probability enters in both places. 

For zero rejected trials, the modified weight distribution is:
\begin{equation}
\frac{d{\cal P}'_0}{dt} = \underbrace{R'_\mrm{acc}(t)}_{\mbox{\tiny reweight}}\underbrace{P_\mrm{acc}(t)}_{\mbox{\tiny accept trial}} \underbrace{\hat{P}(t)\  \hat{\Delta}(t_0,t)}_{\mbox{\tiny generate trial}} = P'(t)\ \hat{\Delta}(t_0,t)~,
\end{equation}
for one rejected trial,
\begin{eqnarray}
\frac{d{\cal P}'_1}{dt} &=& \underbrace{R'_\mrm{acc}(t)}_{\mbox{\tiny reweight}}\underbrace{P_\mrm{acc}(t)}_{\mbox{\tiny accept trial}} \hat{P}(t) \hat{\Delta}(t_0,t) \int_{t}^{t_0} dt_1 \underbrace{R'_\reject(t)}_{\mbox{\tiny reweighting}} \underbrace{\left(\hat{P}(t_1)-P(t_1)\right)}_{\mbox{\tiny reject trial}} \\
&=& P'(t) \ \hat{\Delta}(t_0,t) \int_{t}^{t_0} dt_1  \left(\hat{P}(t_1)-P'(t_1)\right)~,
\end{eqnarray}
and for two rejected trials,
\begin{equation}
\frac{d{\cal P}'_2}{dt} 
 = P'(t) \ \hat{\Delta}(t_0,t) \int_{t}^{t_0} dt_1  \left(\hat{P}(t_1)-P'(t_1)\right) \int_{t}^{t_1} dt_2 
 \left(\hat{P}(t_2)-P'(t_2)\right)~,\label{eq:P2prime}
\end{equation}
hence exactly the same structure emerges for the reweighted sample as for the underlying veto algorithm above, just with $P$ replaced by $P'$. The proof that  \eqRef{eq:resummedPrime} results from the sum over all possibilities is therefore identical to the proof of the original (unweighted) veto algorithm above.  

Two remarks are in order. Firstly, we emphasize that the relative reject-ratio, \eqRef{eq:Rreject}, contains the difference $\hat{P}-P$ in the denominator. This means that, if the trial overestimate, $\hat{P}$, is ``too perfect'' (meaning it is very close to $P$), the denominator can become close to singular, resulting in large and possibly numerically unstable weights.  Algorithmically, what happens is that there are very few failed trials, hence the modifications to the Sudakov factor are not mapped out very well; each failed trial will have a very large job to do. Technically, we address this by applying a ``headroom factor'' to the trial functions when automated uncertainty-variations are requested, ensuring that there is always a non-negligible probability for trials to be rejected at the cost of computational
speed. By default, we choose a headroom factor of 2. For the representative example of hadronic $Z$ decays, this results in a slowdown of the code of only about 20\%.

Secondly, the final event weight, $w'$, after the full shower evolution, is the product of many such factors, one $R'_\mrm{acc}$ for each accepted trial and one $R'_\reject$ for each rejected one,
\begin{equation}
  w' ~= \prod_{i\in \mrm{accepted}}\frac{P'_{i,\mrm{acc}}}{P_{i,\mrm{acc}}} \prod_{j\in \mrm{rejected}} \frac{P'_{j,\reject}}{P_{j,\reject}}~.
\end{equation}
Given enough phase space for evolution, this factor can become arbitrarily different from unity, representing that, e.g., a very active shower history is exponentially more likely to occur in a shower with a large value of $\alpha_s$ than in one with a small  value. In principle, this \emph{is} both physically and mathematically correct. In practice, however, it is not desirable that branchings at low evolution scales in the shower  should significantly alter the modified event weights. Technically, we treat this by imposing a few limiting factors on the variations, as detailed below. 

\subsection{Veto Algorithm with Biased Kernels \label{sec:bias}}

A second important use case for shower algorithms is to evaluate the fragmentation contributions to processes like photon and $B$ hadron production, via splittings like $q\to q\gamma$ and $g\to b\bar{b}$ respectively. ($\pi^0\to\gamma\gamma$ and similar hadron decay processes obviously contribute substantially to the former as well; our focus here is on the perturbative contributions only.) Since these processes are relatively rare ($\alpha_\mrm{em} \ll \alpha_s$ and $P_{g\to b\bar{b}} \ll P_{g\to gg}$), the generation of adequate event samples featuring these processes can suffer from substantial inefficiencies. A complementary case is the generation of high-multiplicity minimum-bias samples in $pp$ collisions, for which events enriched in the number of perturbative MPI could help to improve the generation efficiency (though of course there is also a contribution from events with few MPI but very active hadronization steps). 

A similar line of argument as above allows us to construct weighted samples enriched in these
processes, while preserving the exact Sudakov factors. We note that this method is formally 
identical to the one presented for $q\to q\gamma$ branchings in Ref.~\cite{Hoeche:2009xc}; we 
include its definition and all-orders proof here mostly for completeness, and to have it 
presented in the same notation as above. 

Consider that we wish to enhance the rate of $g\to b\bar{b}$ splittings by a factor of 10 (for 
example), until we have obtained at least one such splitting, after which we would normally 
want to let the probability to have a second $g\to b\bar{b}$ splitting in the same event drop 
back down to the normal level.  We can achieve this by first increasing  the rate of trials for the corresponding splitting function by a factor of 10 by using a larger (biased) trial function (suppressing the dependence on both $t$ and $z$),
\begin{equation}
\hat{P}_\mrm{biased} = 10 \hat{P}~.
\end{equation}
We then keep the accept probability the same as normal, but reweight each accepted biased trial branching by the inverse of the biasing factor, 
\begin{equation}
P_\mrm{acc} ~=~\frac{P}{\hat{P}} ~~~;~~~R_\mrm{acc}~=~\frac{\hat{P}}{\hat{P}_\mrm{biased}}~=~\frac{1}{10}~,\label{eq:biasWeight}
\end{equation}
so that the product $R_\mrm{acc}P_\mrm{acc} \hat{P}_\mrm{biased} = P$ is the desired physical distribution. For each rejected biased trial branching, we use the same technology as above to reweight the event,
\begin{equation}
R_\reject~=~\frac{1-P_\mrm{acc}R_\mrm{acc}}{1-P_\mrm{acc}}~=~ 
\frac{\hat{P}}{\hat{P} - P}\left(1 - \frac{P}{\hat{P}_\mrm{biased}} \right) ~\stackrel{P\ll \hat{P}_\mrm{biased}}{\to}~ \frac{\hat{P}}{\hat{P} - P}~,
\label{eq:biasSudakov}
\end{equation}
where the last asymptotic shows that the reweighting factor becomes independent of the bias in the limit that the bias factor is very large. Nonetheless, the difference is important since, as we shall see below, this is what allows us to recover the physical Sudakov factor. 

We note that if one is interested only in enhancing a single branching of the given type, all events featuring the branching will be accompanied by a single power of the constant inverse-bias factor, \eqRef{eq:biasWeight}, hence that weight can alternatively just be applied to the event sample as a whole, and will cancel in any normalized distributions. The important part is thus the application of \eqRef{eq:biasSudakov} to each rejected trial branching, in order to recover the physical Sudakov factor. Similarly to above, this is a procedure that will only work well when there is at least a minimal number of rejected trial branchings, ensured e.g., by choosing $\hat{P} > 1.2 P$. 

The proof is as follows. For zero rejected trials, the distribution obtained by the above procedure is:
\begin{equation}
\frac{d{\cal P}_0}{dt} = \underbrace{R_\mrm{acc}(t)}_{\mbox{\tiny reweight}}\underbrace{P_\mrm{acc}(t)}_{\mbox{\tiny accept trial}} \underbrace{\hat{P}_\mrm{biased}(t)\  \hat{\Delta}_\mrm{biased}(t_0,t)}_{\mbox{\tiny generate biased trial}} = P(t)\ \hat{\Delta}_\mrm{biased}(t_0,t)~,
\end{equation}
for one rejected trial,
\begin{eqnarray}
\frac{d{\cal P}_1}{dt} &=& P(t) \ \hat{\Delta}_\mrm{biased}(t_0,t) \int_{t}^{t_0} dt_1 \hat{P}_\mrm{biased}(t_1)\left(1-P_\mrm{acc}(t_1)\right)
R_\reject(t_1)  \\
&=& P(t) \ \hat{\Delta}_\mrm{biased}(t_0,t) \int_{t}^{t_0} dt_1  \left(\hat{P}_\mrm{biased}(t_1)-P(t_1)\right)~,
\end{eqnarray}
where the second equality follows from the exact definition of $R_\reject$ in \eqRef{eq:biasSudakov}, while the asymptotic version of it would only generate the first term in the integrand. For two rejected trials,
\begin{equation}
\frac{d{\cal P}_2}{dt} 
 = P(t) \ \hat{\Delta}_\mrm{biased}(t_0,t) \int_{t}^{t_0} dt_1  \left(\hat{P}_\mrm{biased}(t_1)-P(t_1)\right) \int_{t}^{t_1} dt_2 
 \left(\hat{P}_\mrm{biased}(t_2)-P(t_2)\right)\label{eq:P2biased}~.
\end{equation}
As required, the nested integrals translate between $\hat{P}_\mrm{biased}$ and the physical branching probability, $P$, such that the produced Sudakov factors will depend only upon $P$, not $P_\mrm{biased}$. 

\subsection{Renormalization-Scale Variations
\label{sec:muRfac}}

The first major class of variations we include are variations of the shower renormalization scales. This can be done for both QED and QCD, with the latter normally dominating the overall uncertainty. It is worth noting, however, that for a coherent shower algorithm, a scale choice of $p_\perp$ accompanied by the so-called CMW scale factor~\cite{Amati:1980ch,Catani:1990rr} absorbs the leading second-order corrections to the splitting functions for soft-gluon emission. A brute-force scale variation would destroy this agreement. We therefore provide an option to allow an explicit ${\cal O}(\alpha_s^2)$ compensating term to accompany each scale variation,  driving the effective scale choice back towards $p_\perp$ at the NLO level, while leaving the higher-order components of the scale variation untouched. 

Specifically, if the baseline gluon-emission density is
\begin{equation}
P(t,z) ~=~ \frac{\alpha_s(p_\perp)}{2\pi} \frac{P(z)}{t} ~,
\end{equation}
with $P(z)$ the DGLAP radiation kernel, then we may define a renormalisation-scale variation, $\mu=p_\perp \to \mu'=kp_\perp$, with an NLO-compensating term (see, e.g.,~\cite{Hartgring:2013jma})
\begin{equation}
P'(t,z) ~=~ \frac{\alpha_s(k p_\perp)}{2\pi}\left(1 + \frac{\alpha_s}{2\pi}\beta_0 \ln k\right) \frac{P(z)}{t} ~,\label{eq:compensation}
\end{equation}
with $\beta_0=(11N_C - 2n_F)/3$, $N_C=3$, and $n_F$ the number of active flavours at the scale $\mu = p_\perp$. 
Note that, if there are any quark-mass thresholds in-between $p_\perp$ and $kp_\perp$, then $\alpha_s(p_\perp)$ and $\alpha_s(kp_\perp)$ will not be evaluated with the same $n_F$. Matching conditions are applied in PYTHIA to make the running continuous across thresholds, so this effect should be small for reasonable values of $k$. Nonetheless one could in principle add an additional term $\alpha_s/(2\pi) \ln(m_q/(kp_\perp)) / 3$ to compensate for the different $\beta_0$ coefficients used in the region between the threshold and $kp_\perp$; however since the variation is numerically larger without that term, and since the ambiguities associated with thresholds are anyway among the uncertainties one could wish to explore, for the time being we consider it more conservative to not include any such terms.

Note also that the scale and scheme of the $\alpha_s$ factor in the compensation term, inside the parenthesis in \eqRef{eq:compensation}, is not specified, as this amounts to an effect of yet higher order, beyond NLO. To make the compensation as conservative as possible (and to avoid the risk of over-compensating), we choose the scale of the compensation term to be the largest local scale in the problem, namely the invariant mass of the emitting colour dipole
$m_\mrm{dip}$, thus making the correction term as numerically small (and hence as conservative) as possible; specifically $\mu_\mrm{max}=\max(m_\mrm{dip},kp_\perp)$. 
Furthermore, since the analyses of~\cite{Amati:1980ch,Catani:1990rr} only pertain to the soft limit, our estimate of the compensation would be too optimistic if applied undiminished over all of phase space. To be more conservative, we therefore multiply the compensation term by an explicit factor $(1-\zeta)$, defined so as to vanish linearly outside the soft limit,
\begin{equation}
\zeta = \left\{ 
\begin{array}{ccl}
z&&\mbox{for splittings with a $1/z$ singularity}\\ 
1-z&&\mbox{for splittings with a $1/(1-z)$ singularity}\\
\min(z,1-z)&&\mbox{for splittings with a $1/(z(1-z))$ singularity}
\end{array}
\right.~.
\end{equation}
Combined, these arguments lead us to the following modified accept probability for a robust shower renormalisation-scale variation compatible with the known second-order leading-singular structure:
\begin{equation}
P'(t,z)~=~ \frac{\alpha_s(k p_\perp)}{2\pi}\left(1 + (1-\zeta)\frac{\alpha_s(\mu_\mrm{max})}{2\pi}\beta_0 \ln k\right)\frac{P(z)}{t}~,
\end{equation}
hence 
\begin{equation}
R'_\mrm{acc}(t,z)~=~\frac{P_\mrm{acc}'(t,z)}{P_\mrm{acc}(t,z)} ~=~ \frac{\alpha_s(k p_\perp)}{\alpha_s(p_\perp)}\left(1 + (1-\zeta)\frac{\alpha_s(\mu_\mrm{max})}{2\pi}\beta_0 \ln k\right)~.
\end{equation}

We emphasize that the compensation term in the expressions above is only included for gluon emissions, not for $g\to q\bar{q}$ splittings. The latter are subjected to the full (uncompensated) variation, $\alpha_s(kp_\perp)/\alpha_s(p_\perp)$.

Finally, we impose an absolute limit on the allowed amount of $\alpha_s$ variation, by default
\begin{equation}
    |\Delta \alpha_s| \le 0.2~.
\end{equation}
This does not significantly restrict the range of variation for perturbative branchings (even when $\alpha_s \sim 0.5$, a full 40\% amount of variation is still allowed), but it does prevent branchings very near the cutoff from generating large changes to the event weights. Removing this bound would not significantly affect the perturbative physics uncertainties, but would cause much larger weight fluctuations (between events with and without some very soft branching near the end of the evolution), mandating much longer run times for the same statistical precision. 

At the technical level, the user decides whether to perform scale variations of ISR and FSR
independently, or whether to vary the respective $\alpha_s$ factors in a correlated manner. It 
is even possible to include both types of variations (independent and correlated), and compare 
the results obtained at the end of the run. From a practical point of view, the FSR $\alpha_s$ 
choice mainly influences the amount of broadening of the jets, while the ISR $\alpha_s$ choice 
influences resummed aspects such as the combined recoil given to a hard system (e.g., a $Z$, 
$W$, or $H$ boson, or a $t\bar{t}$, dijet, or $\gamma+\mrm{jet}$ system) by ISR radiation and 
also how many extra jets are created from ISR. The latter of course also depends on whether and
how corrections from higher-order matrix elements are being accounted for. 

\begin{figure}[t]
\centering
\includegraphics*[scale=0.385]{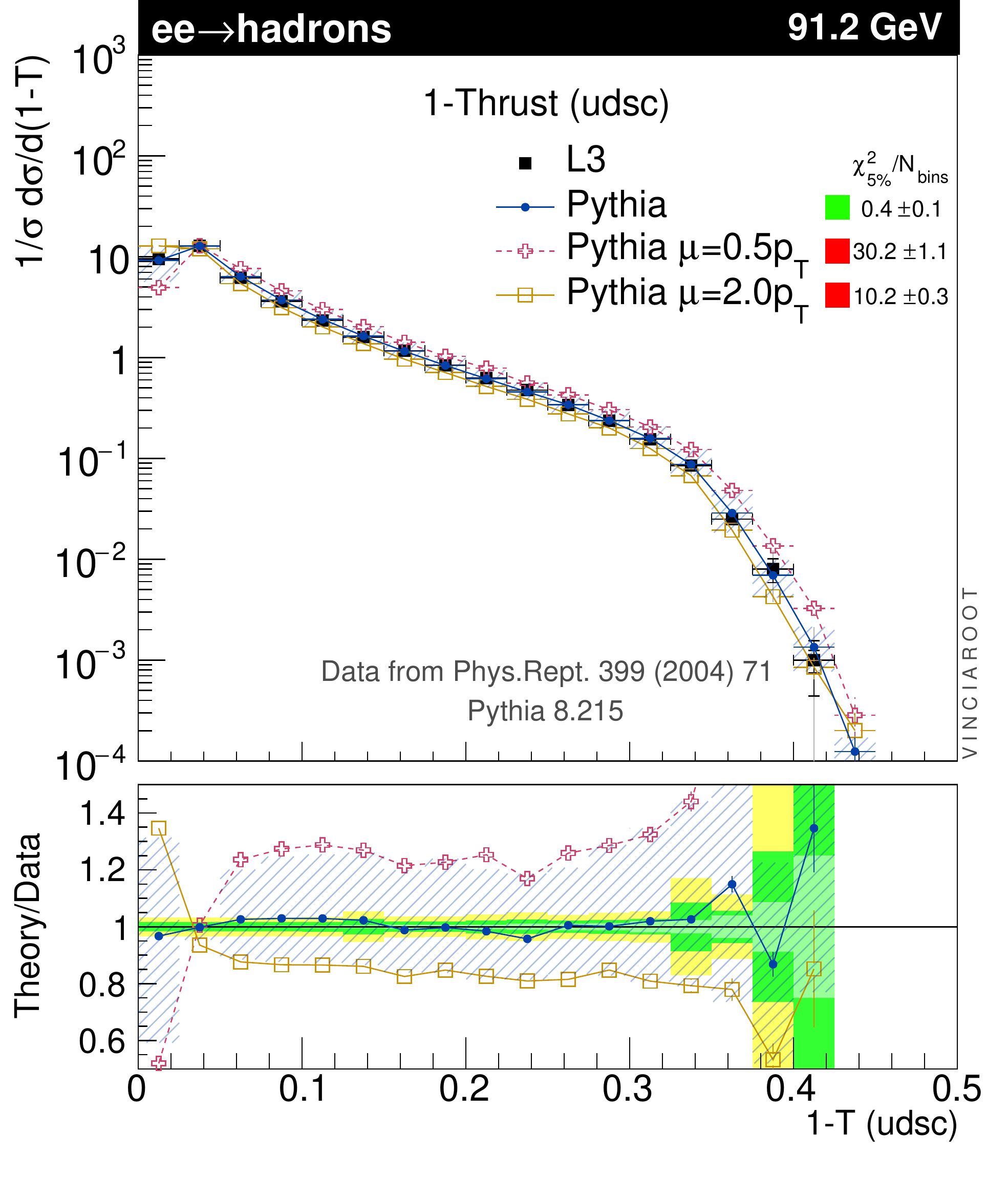}
\includegraphics*[scale=0.385]{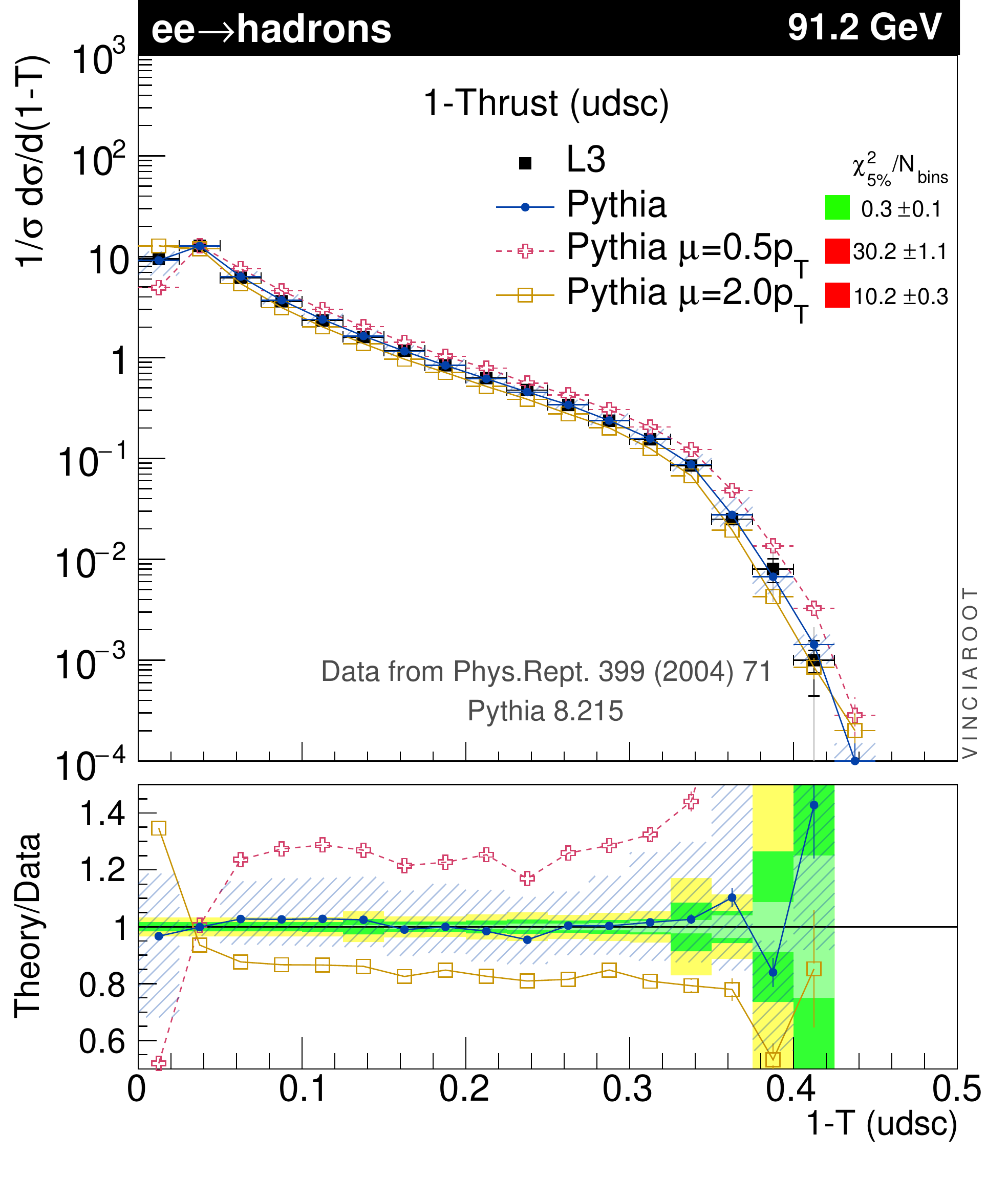}
\caption{Illustration of the default renormalisation-scale variations for FSR, by a factor of 2 in each direction. The central (default, unweighted) shower calculation is shown in blue, with $///$ hashing indicating the range spanned by the variation weights. The dashed (red) and solid (yellow) lines represent the results of standalone runs with $\mu_R=0.5p_\perp$ and $\mu_R=2p_\perp$ respectively. {\sl Left:} without the NLO scale-compensation term. {\sl Right:} with the NLO scale-compensation term (the default setting). Distribution of 1-Thrust for $e^+e^-\to\mrm{hadrons}$ at the $Z$ pole, excluding $b$-tagged events; ISR switched off;  data from the L3 experiment~\cite{Achard:2004sv}.
\label{fig:muFSR}}
\end{figure}
An illustration and validation of the automated renormalisation-scale variations is given in
\figRef{fig:muFSR}, for the case of FSR and the distribution of 1-Thrust in
$e^+e^-\to\mrm{hadrons}$ events at the $Z$ pole, compared to a measurement by the L3
experiment~\cite{Achard:2004sv}. (QED ISR is switched off and $b$-tagged events are excluded
in this comparison.) First, we perform three separate dedicated runs, using $\mu_R = 2p_\perp$ (solid yellow lines with square symbols), $\mu_R = p_\perp$ (the default choice, solid blue lines with dot symbols), and $\mu_R = 0.5p_\perp$ (dashed red lines with open $+$ symbols). For the central run, we also included the automated weight variations presented here, for the same factor-2 $\mu_R$ variations. The range spanned by the reweighted central distribution is shown by the blue $///$ hashed areas. On the left-hand side of \figRef{fig:muFSR}, the NLO scale-compensation term is switched off, and we see that the results of the independent runs are faithfully reproduced by the reweighted central-run distributions. (The small difference in the first bin is due to the absolute limit of $|\Delta\alpha_s| \le 0.2$ which we impose in the reweighting framework.) On the right-hand side of \figRef{fig:muFSR}, the same distributions are shown, but now with the NLO scale-compensation term switched on. The difference between the standalone runs (where no compensation is applied) and the reweighted distributions illustrates the effect of the compensation term.

\begin{figure}[tp]
\centering
\includegraphics*[scale=0.385]{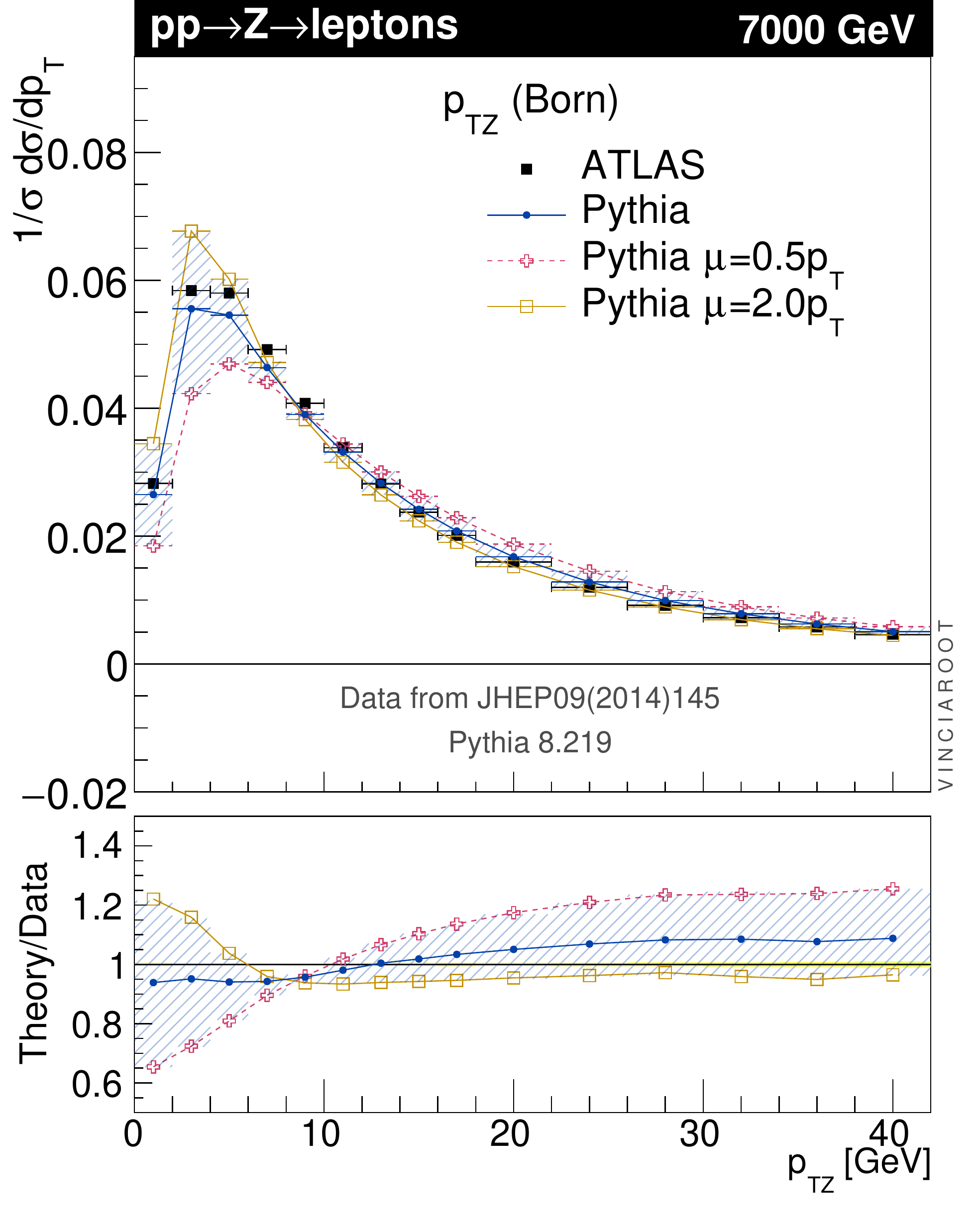}
\includegraphics*[scale=0.385]{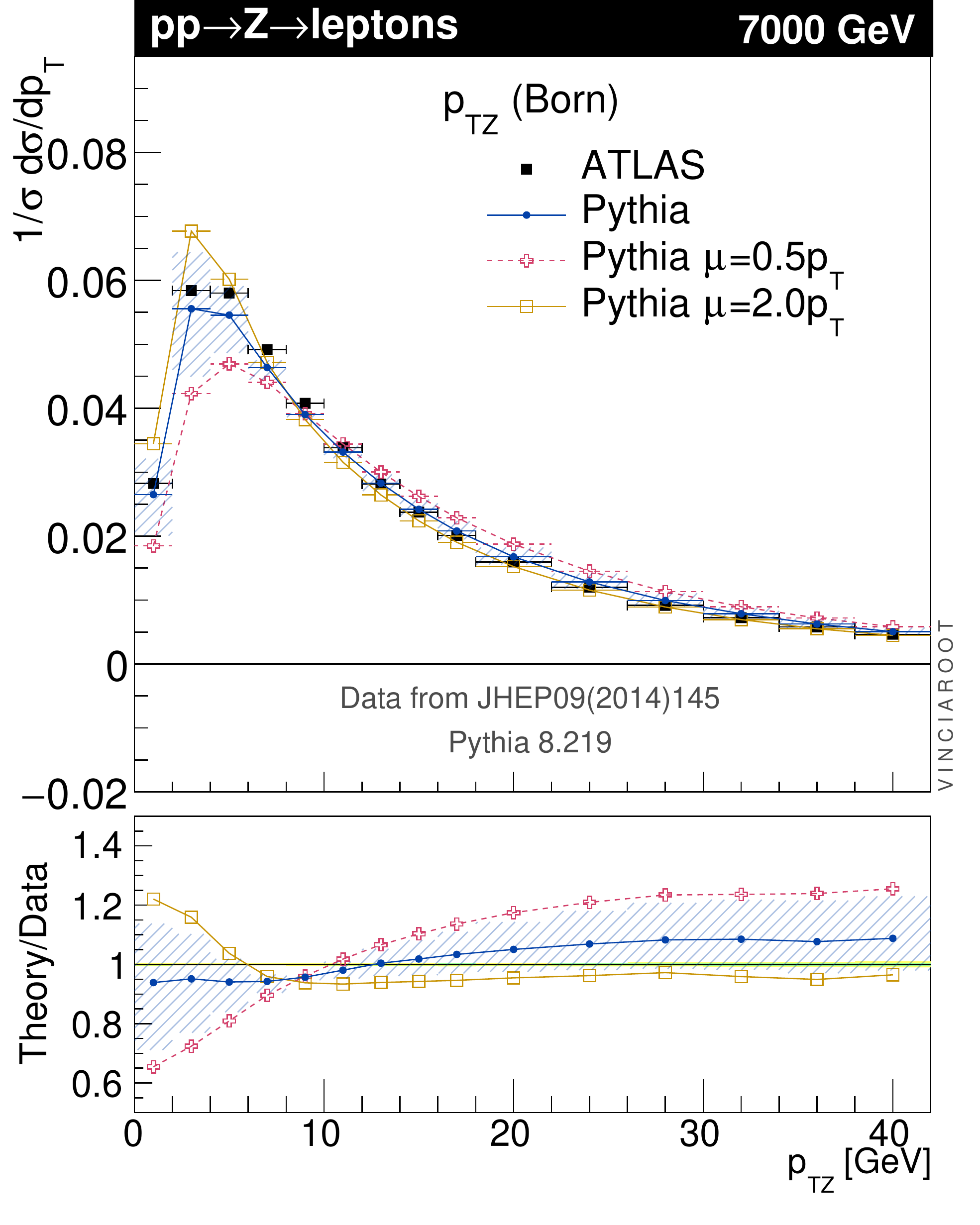}\\
\includegraphics*[scale=0.385]{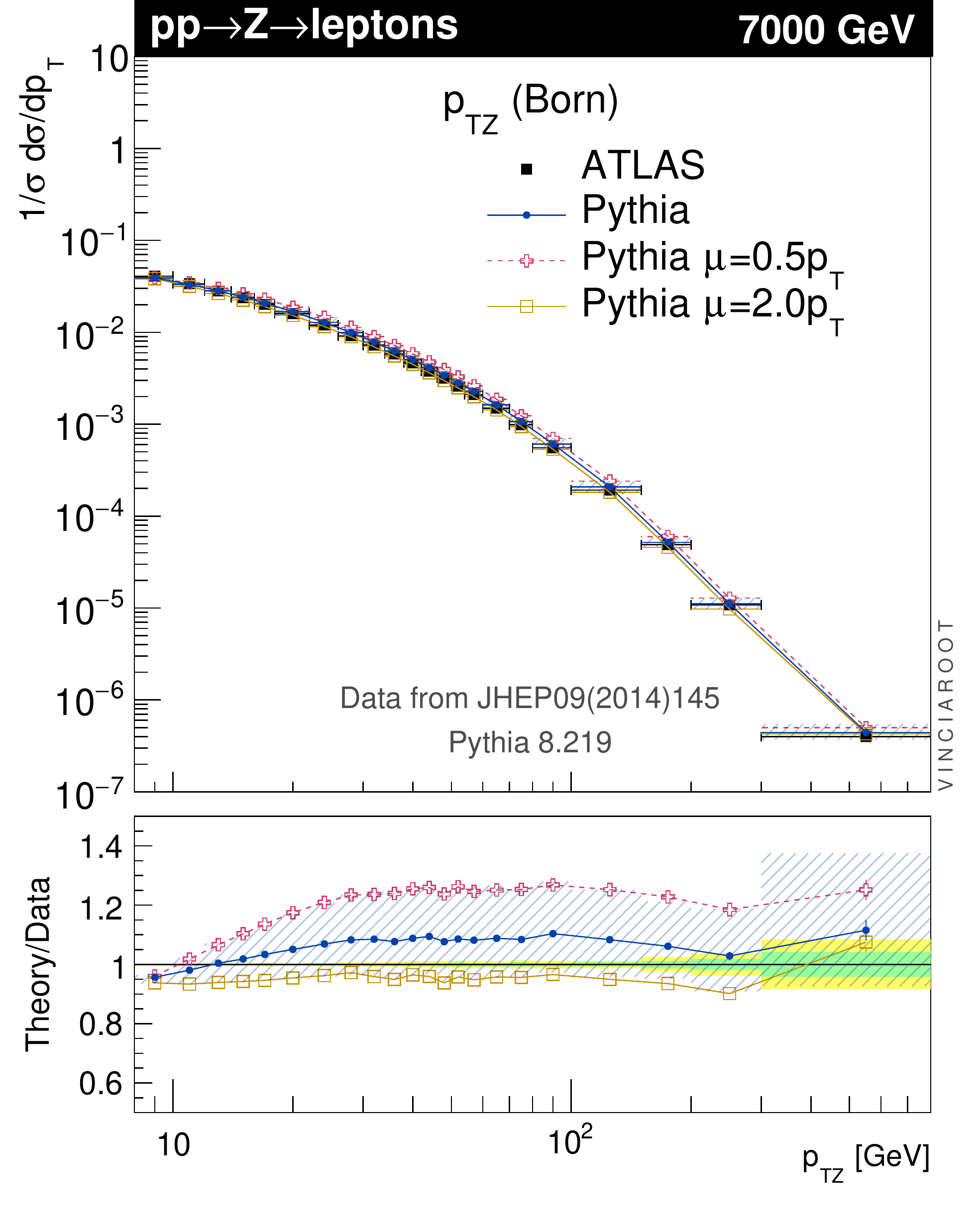}
\includegraphics*[scale=0.385]{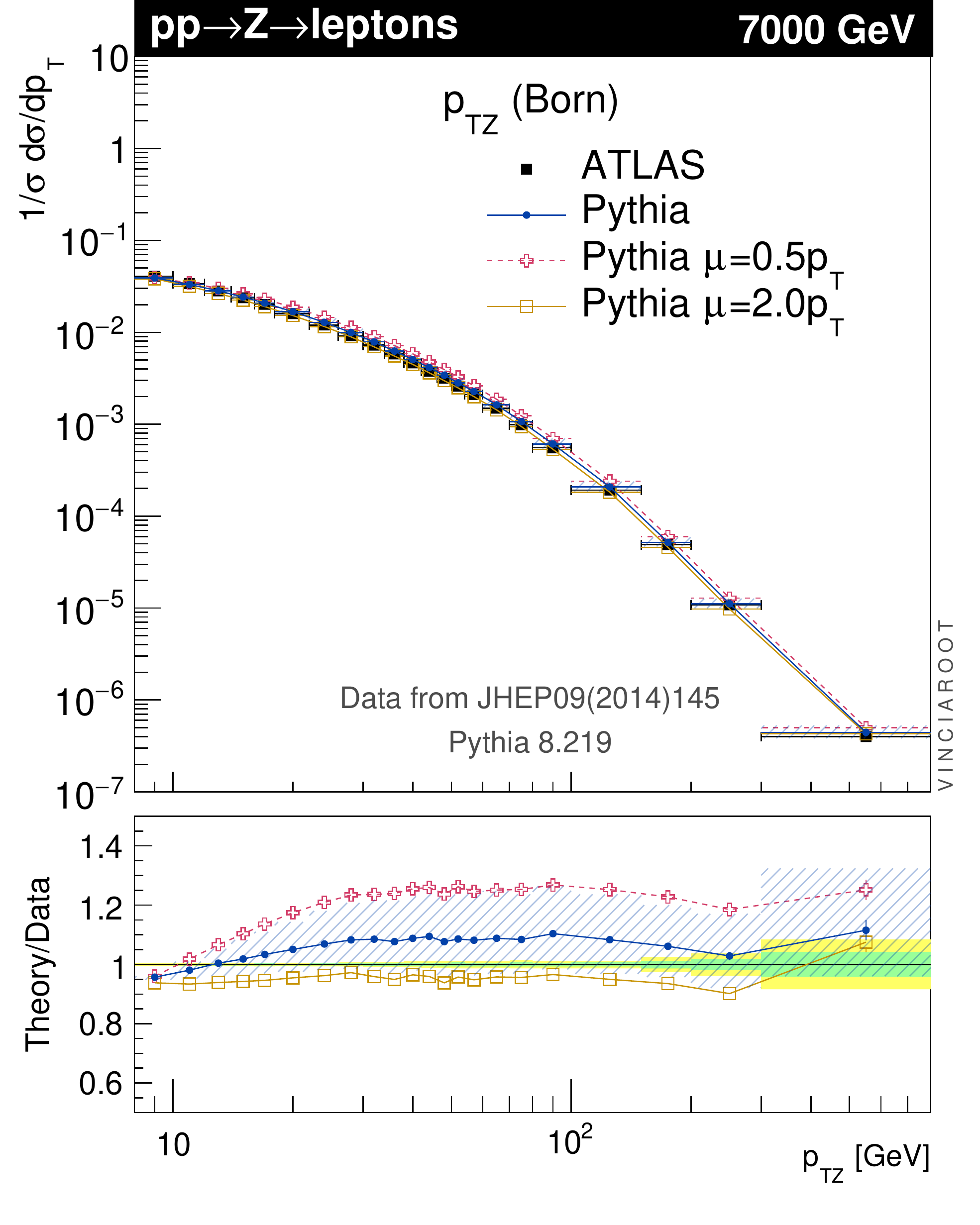}
\caption{Illustration of the default renormalisation-scale variations for ISR, by a factor of 2 in each direction. The central (default, unweighted) shower calculation is shown in blue, with $///$ hashing indicating the range spanned by the variation weights. The dashed (red) and solid (yellow) lines represent the results of standalone runs with $\mu_R=0.5p_\perp$ and $\mu_R=2p_\perp$ respectively. {\sl Left:} without the NLO scale-compensation term. {\sl Right:} with the NLO scale-compensation term (the default setting). Distribution of the $p_\perp$ spectrum of the lepton pair in $pp\to Z\to e^+e^-/\mu^+\mu^-$ at the $Z$ pole ($66<m_{\ell\ell}/\mrm{GeV}<116$), for leptons in the phase-space window $|\eta_\ell|<2.4$, $p_{\perp\ell}>20$ GeV;  data from the ATLAS experiment~\cite{Aad:2014xaa}.
\label{fig:muISR}}
\end{figure}
A corresponding validation for the initial-state shower renormalisation-scale variations is given in \figRef{fig:muISR}, where we have chosen the transverse momentum of the lepton pair in Drell-Yan events as the test observable. The peak region below $p_{\perp Z}=40~\mrm{GeV}$ is shown in the top row of plots (on a linear scale) while the bottom row shows the tail of the spectrum (on a log-log scale). As in \figRef{fig:muFSR}, the hashed area in the plots in the left-hand (right-hand) column shows the uncertainty band with the NLO scale-compensation term switched off (on). The effect is here less than in the FSR case, cf.~\figRef{fig:muFSR}, presumably due to the compensation term being proportional to $\alpha_s(m_\mrm{dip})$ where $m_\mrm{dip}$ can be very large in the ISR case. 

Note to experimentalists: rather than performing dedicated runs for $\mu_R$ variations, we recommend using the uncertainty weights instead, since the renormalisation-scale compensation term is only available for the latter and allows slightly more aggressive (smaller) uncertainty estimates.

\subsection{Splitting-Kernel Variations \label{sec:cNS}}

All shower formalisms are based upon the universal nature of the singular infrared (soft and/or collinear) limits of QCD. In these limits, the exact form of the splitting functions are known (to a given order), regardless of whether we express them as DGLAP kernels, dipole/antenna functions, or by any other means. Away from these limits, however, in the physical phase space on which the kernels will be applied as approximations, there are in principle infinitely many different radiation functions to choose from, sharing the  same singular terms but having different nonsingular ones. This represents a fundamental ambiguity for shower algorithms which cannot be evaded by, e.g., setting the non-singular terms to zero. Firstly, any such (arbitrary) choice would not address the underlying issue. Secondly, it would not be stable against reparametrisations of the radiation functions themselves. For example, zero in one dipole parametrisation does not correspond to zero in another, see~e.g.~\cite{Giele:2011cb,LopezVillarejo:2011ap}. 

Moreover, varying the splitting kernels by nonsingular (a.k.a.\ ``finite'') terms produces uncertainty envelopes which are quite complementary to those produced by renormalisation-scale variations~\cite{Giele:2011cb}. The reason is that renormalisation-scale variations are by construction proportional to the (default) shower radiation functions, while nonsingular terms vary the radiation functions themselves. In regions far from the singular limits, the pole terms are highly suppressed and the default shower radiation functions may not bear much resemblance to the matrix elements for the process at hand. In such regions, process-dependent  nonsingular terms dominate, and corresponding nonsingular-term variations in the shower radiation functions can therefore easily produce much larger (and more realistic) uncertainty estimates than renormalisation-scale changes. 

We therefore believe that an exhaustive exploration of parton-shower uncertainties should at least grant the \emph{capability} to perform nonsingular variations of the shower kernels, while the final decision whether and how to use them  can still be left up to the user.  
An observation of large nonsingular-term uncertainties in the context of a physics study would be a direct indication of a need to incorporate further corrections from matrix elements, e.g.\ via one of the many matching/merging strategies available in PYTHIA 8. This is because the matrix elements contain the correct (process-dependent) nonsingular terms for the  process at hand, thus nullifying the nonsingular-term uncertainties at least in any phase-space regions populated by the matrix elements. 

To implement such variations in the context of a DGLAP approach, we allow for the following modification of the shower splitting kernels,
\begin{equation}
\frac{P(z)}{Q^2} \ dQ^2 \ \to\  \left(\frac{P(z)}{Q^2} + \frac{\cns}{m_\mrm{dip}^2}\right) dQ^2 = \left(P(z) + \frac{\cns \ Q^2}{m_\mrm{dip}^2}\right) \frac{dt}{t}~,
\end{equation}
where $m_\mrm{dip}$ is the invariant mass of the dipole in which the splitting occurs, $\cns$ is a dimensionless constant of order unity which parametrises the amount of (nonsingular) splitting-kernel variation, and in the last equality we used the identity $dQ^2/Q^2 = dt/t$ which holds for any $t=f(z)Q^2$, including in particular all the PYTHIA evolution variables. Note that, for gluon emission off timelike massive quarks, $Q^2$ should be the virtuality, or off-shellness of the massive quark, defined as $Q^2 = (p_b + p_g)^2 - m_b^2 = 2p_b\cdot p_g$~\cite{Norrbin:2000uu}, with $p_b$ the 4-momentum of the massive quark and $p_g$ that of the emitted gluon. (For spacelike virtual massive quarks, the mass correction has the opposite sign~\cite{Sjostrand:2004ef}.)
Thus,
\begin{equation}
P'(t,z)~=~\frac{\alpha_s}{2\pi}\ {\cal C}\left(\frac{P(z) \ + \ \cns \ Q^2/m^2_\mrm{dip}}{t}\right)~,
\end{equation}
where $\cal C$ is the colour factor. The variation can therefore be obtained by introducing a spurious term proportional to $Q^2/m_\mrm{dip}^2$ in the splitting kernel used to compute the accept probability, hence
\begin{equation}
R'_\mrm{acc}~=~\frac{P'_{\mrm{acc}}}{P_\mrm{acc}}~=~1 + \frac{\cns \ Q^2/m_\mrm{dip}^2}{P(z)}~, 
\end{equation}
from which we also immediately confirm that the relative variation explicitly vanishes when $Q^2\to 0$ or $P(z)\to\infty$. 

To motivate a reasonable range of variations, we take the nonsingular terms that different physical
matrix elements exhibit as a first indicator, and supplement that by considering the terms that
are induced by PYTHIA's matrix-element corrections (MECs) for $Z$ boson
decays~\cite{Bengtsson:1986hr}. In particular, the study in \cite{LopezVillarejo:2011ap} found
order-unity differences (in dimensionless units) between different physical processes and three different antenna-shower formalisms: Lund dipoles a la ARIADNE~\cite{Gustafson:1987rq,Lonnblad:1992tz}, GGG antennae a la VINCIA~\cite{GehrmannDeRidder:2005cm,Giele:2011cb,Larkoski:2013yi}, and Sector antennae a la Kosower~\cite{Kosower:1997zr,LopezVillarejo:2011ap}. Therefore, here we also take variations of
order unity as the baseline for our recommendations. 

\begin{figure}[t]
\centering
\includegraphics*[scale=0.385]{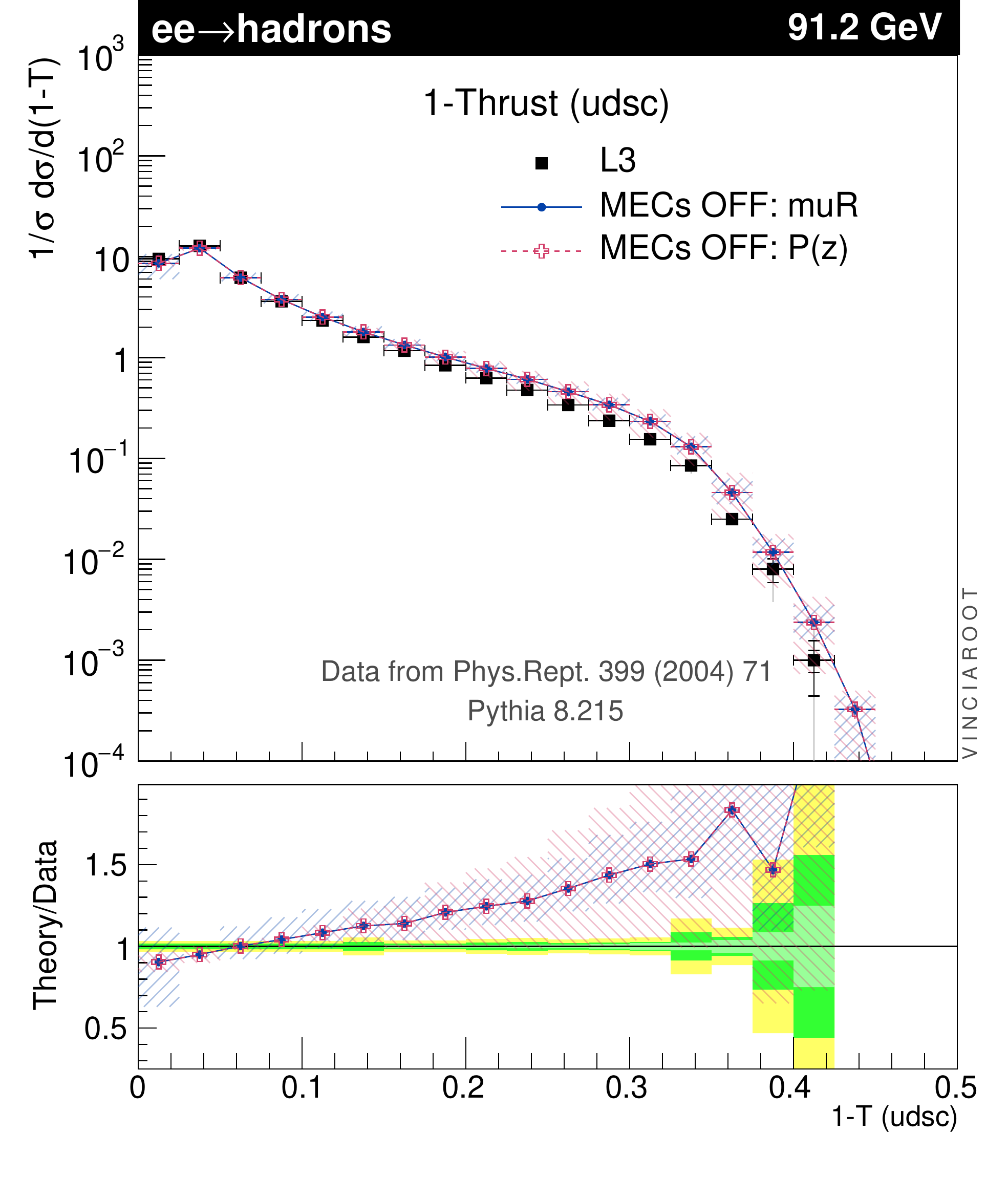}
\includegraphics*[scale=0.385]{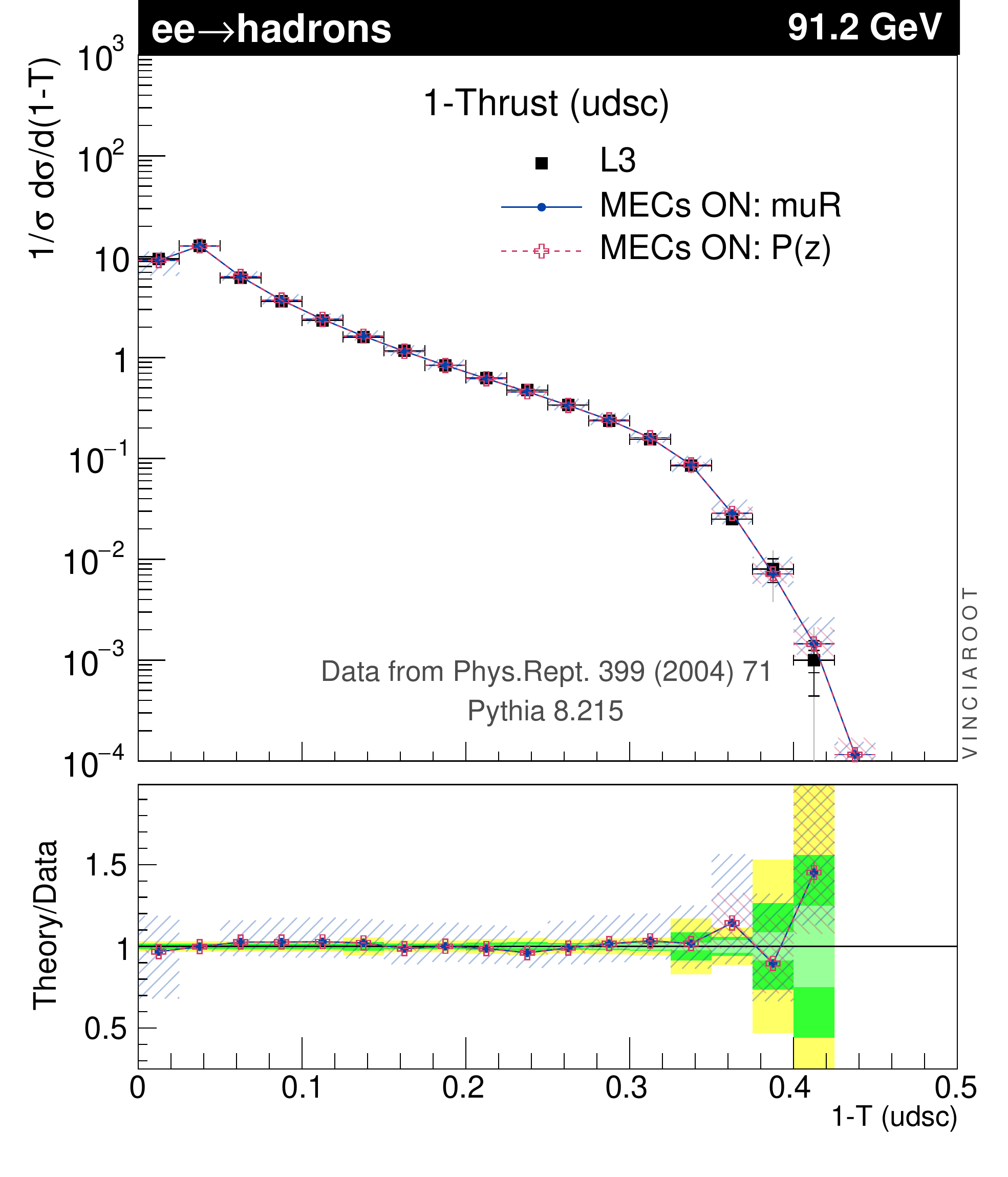}
\caption{Illustration of the default nonsingular variations for FSR splitting kernels, corresponding to $\cns =\pm2$ (shown in red with $\backslash\backslash\backslash$ hashing), compared with the default renormalisation-scale variations by a factor of 2 with the NLO compensation term switched on (shown in blue with $///$ hashing). {\sl Left:} matrix-element corrections OFF. {\sl Right:} matrix-element corrections ON. Note that the range of the ratio plot is greater than in \figRef{fig:muFSR} Distribution of 1-Thrust for $e^+e^-\to\mrm{hadrons}$ at the $Z$ pole, excluding $b$-tagged events; ISR switched off;  data from the L3 experiment~\cite{Achard:2004sv}.
\label{fig:cFSR}}
\end{figure}
In \figRef{fig:cFSR}, we illustrate the splitting-kernel variation taking $\cns = \pm2$ as a first
guess at a reasonable range of variation. As can be observed by comparing the left- and
right-hand panes of the figure, where PYTHIA's MECs are switched off and on respectively, this
variation, labeled $P(z)$ and shown in red with $\backslash\backslash\backslash$ hashing, roughly spans the range between PYTHIA with and without matrix-element corrections. In the right-hand pane, where PYTHIA's internal MECs for $Z\to 3\ \mbox{jets}$~\cite{Bengtsson:1986hr} are switched on, the splitting-kernel uncertainty is essentially zero in the 3-jet region $1-T\le 0.33$, since the nonsingular terms are there provided by the matrix elements. There are in principle still nonsingular-term uncertainties starting from the 4-jet level, beyond $0.33$. Note that the ratio panes in \figRef{fig:cFSR} have a larger range than those of \figRef{fig:muFSR} and that, for comparison, the renormalisation-scale uncertainty, with the scale-compensation term switched on, is still shown in blue with $///$ hashing. 

\begin{figure}[tp]
\centering
\includegraphics*[scale=0.385]{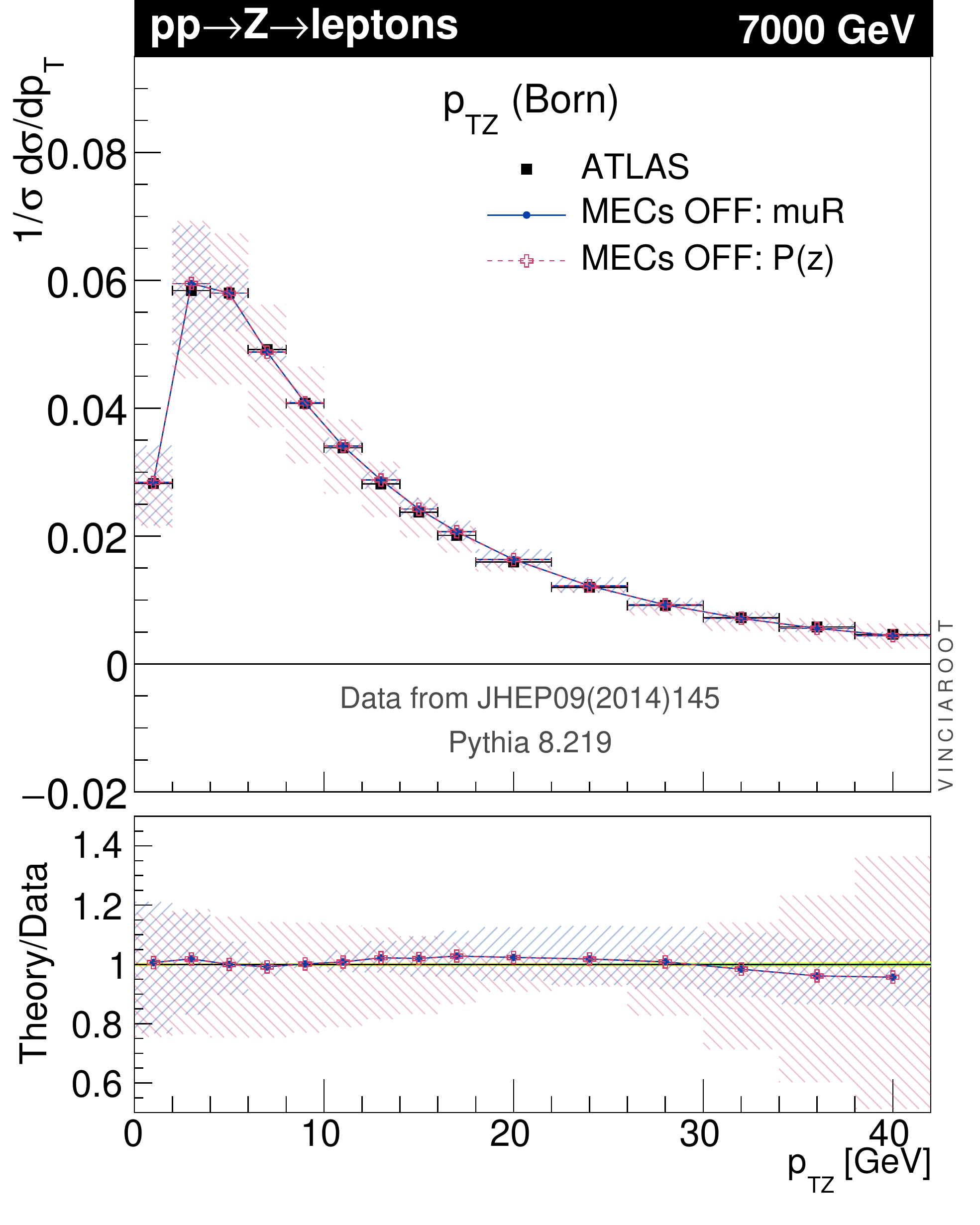}
\includegraphics*[scale=0.385]{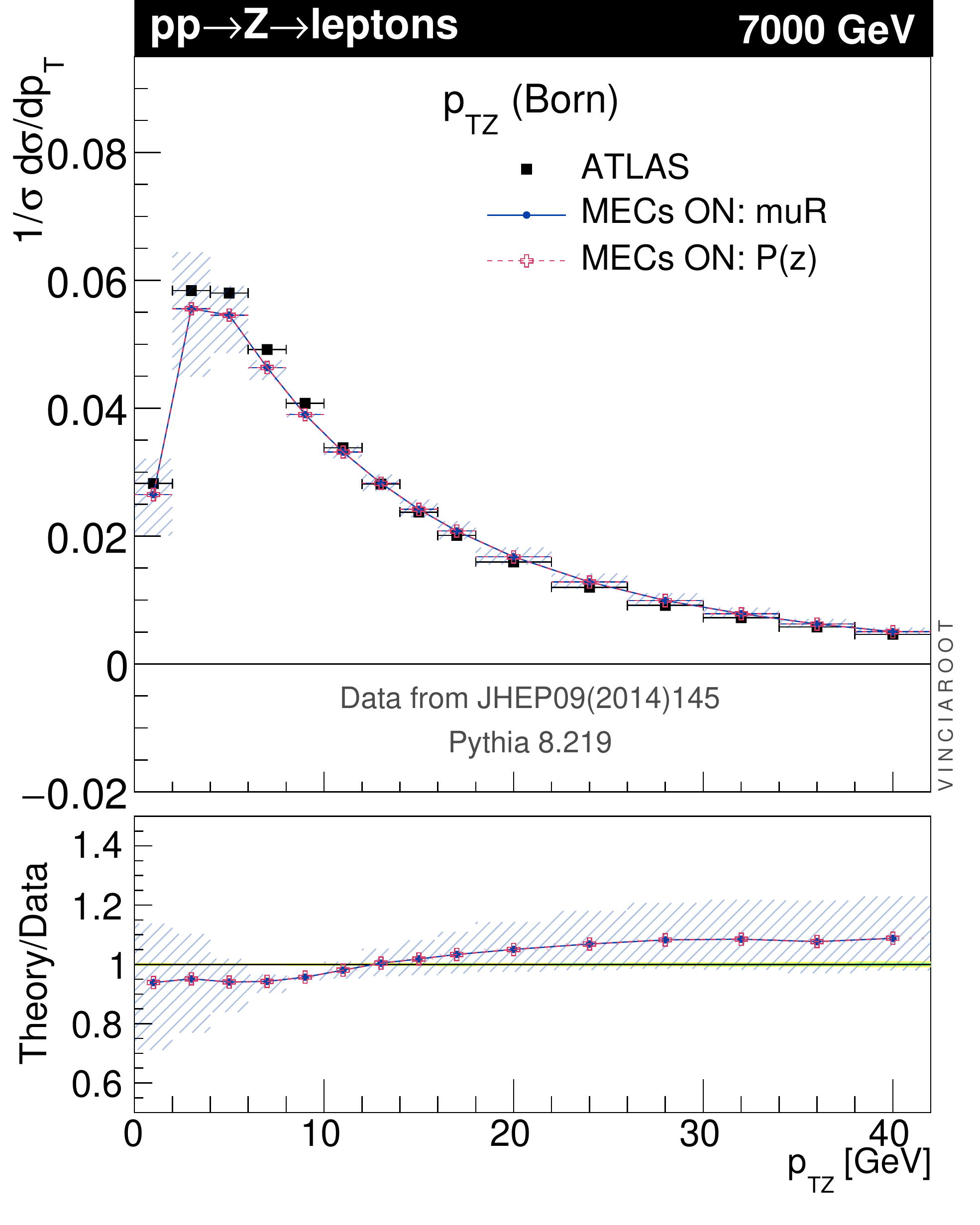}\\
\includegraphics*[scale=0.385]{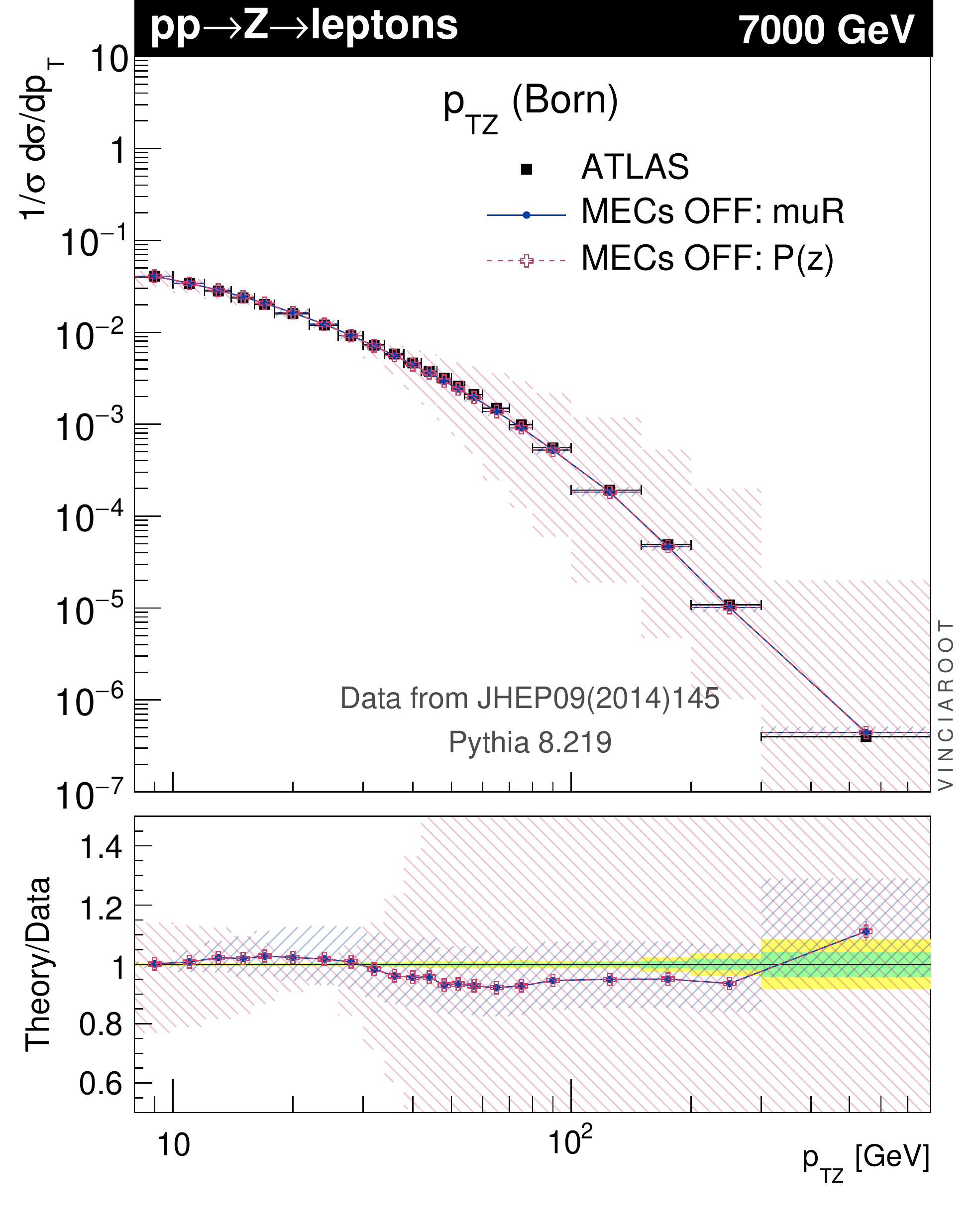}
\includegraphics*[scale=0.385]{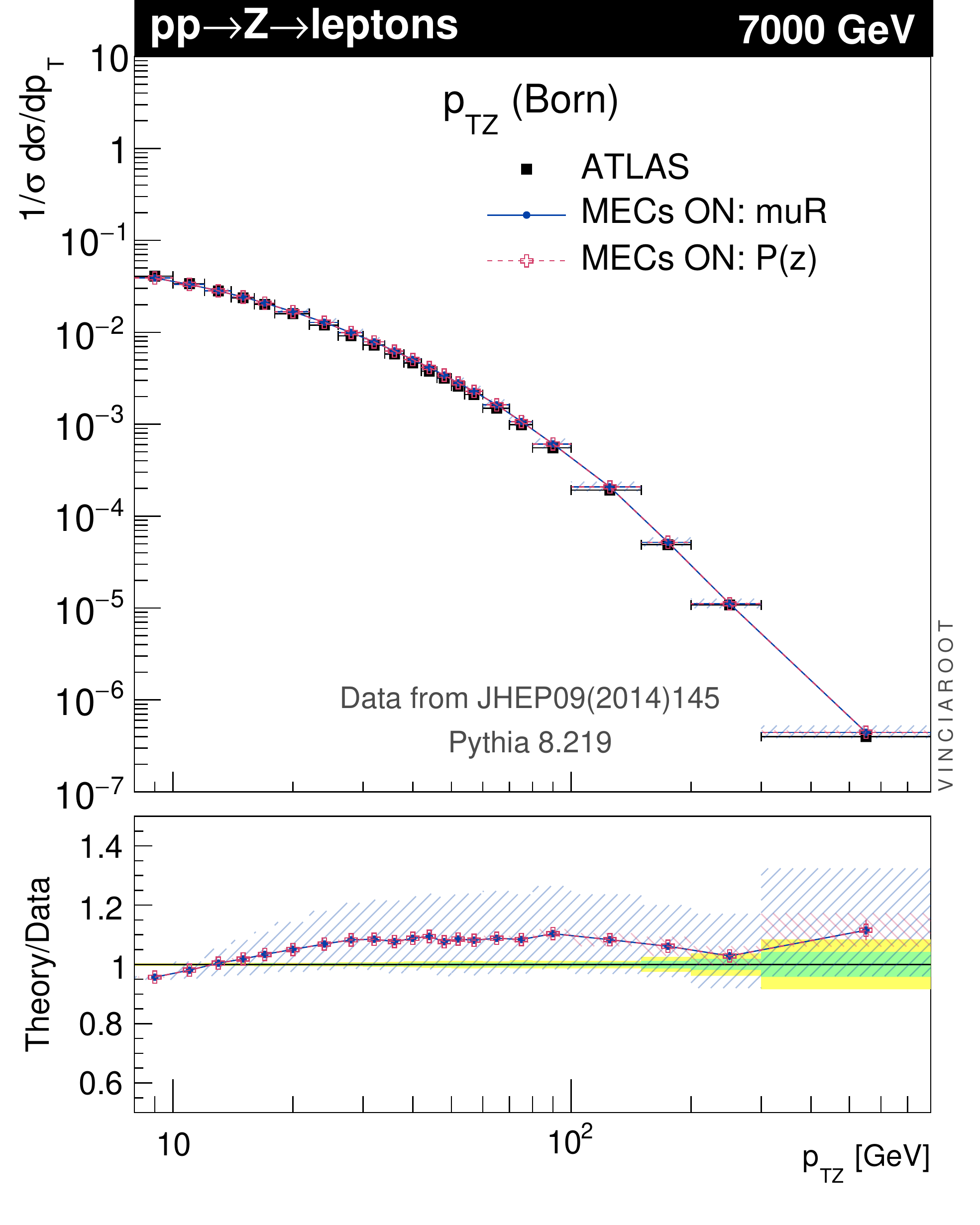}
\caption{Illustration of the default nonsingular variations for ISR splitting kernels, corresponding to $\cns =\pm2$ (shown in red with $\backslash\backslash\backslash$ hashing), compared with the default renormalisation-scale variations by a factor of 2 with the NLO compensation term switched on (shown in blue with $///$ hashing).  {\sl Left:} matrix-element corrections OFF. {\sl Right:} matrix-element corrections ON. Distribution of the $p_\perp$ spectrum of the lepton pair in $pp\to Z\to e^+e^-/\mu^+\mu^-$ at the $Z$ pole ($66<m_{\ell\ell}/\mrm{GeV}<116$), for leptons in the phase-space window $|\eta_\ell|<2.4$, $p_{\perp\ell}>20$ GeV;  data from the ATLAS experiment~\cite{Aad:2014xaa}.
\label{fig:cISR}}
\end{figure}
The case of nonsingular-term variations for the ISR splitting kernels is shown in \figRef{fig:cISR}, again compared to the renormalisation-scale variations (with the NLO compensation term switched on), for the same $p_{\perp Z}$ distributions as were shown in \figRef{fig:muISR}. For this specific case, PYTHIA's matrix-element corrections~\cite{Miu:1998ju} do not have as dramatic an effect on the central prediction as they did for FSR, as can be seen by comparing the central lines of the plots in the left-hand column of the figure (MECs OFF) to the ones on the right (MECs ON). The variation of nonsingular terms, however, is completely cancelled when MECs are switched on, as expected for a distribution dominated by a single emission. 

For completeness, we remark that the reweighting strategies presented here, and parton showers in general, are based on exact cancellation between real and virtual corrections. This is called \emph{detailed balance} and is also referred to as \emph{unitarity} in the parton-shower context. However, the KLN theorem~\cite{Kinoshita:1962ur,Lee:1964is} allows for violations of this balance by non-singular terms. Hence a realistic assessment of the full uncertainties of parton-shower calculations should take into account that non-singular terms can contribute not only in the radiation functions, as above, but also at the level of breaking detailed balance. This would amount to an estimate of the possible size of NLO (and higher) K-factors. To accomplish this consistently, however, several further aspects would need to be addressed, including variations already at the Born level and ensuring that weight modifications at the $n$-th branching in the shower don't change the total cross-section by more than factors proportional to $\alpha_s^{\mrm{Born}+n}$. We deem these considerations to be beyond the scope of this work, but emphasize that they should be investigated.

\section{Summary}

We have described the mathematical formalism and practical validation for a new way of calculating perturbative uncertainty estimates in the PYTHIA 8 Monte Carlo event generator, following the proposal made in~\cite{Giele:2011cb}. Instead of performing independent Monte Carlo runs for each (set of) parameter variation(s), we effectively recycle the vetoed trials  of the Sudakov veto algorithm to provide a numerical mapping of the probability-density changes resulting from different choices of renormalization scales and non-singular terms. The result is cast as a vector of weights for each event whose zero element corresponds to the nominal (user) settings, with the uncertainty variations telling how much the probability to obtain that event would have changed under different showering assumptions. 

Each set of weights is separately unitary, in the sense that they integrate to the same total inclusive cross section for the process at hand. It is therefore important to note that non-unitary changes, such as ``K-factor'' variations, are not accessed by this framework, but would have to be estimated separately. The same is true for PDF variations and for variations of the non-perturbative fragmentation parameters.

The variation weights can be interpreted as follows: branching sequences dominated by well-controlled logarithmically enhanced splittings will produce small weight variations, while events containing one or more branchings for which PYTHIA's underlying assumptions may be compromised will exhibit large weight variations.  
Large non-singular-term uncertainties should be taken as indicating a need for including more matched matrix elements in the calculation, since even LO matrix elements contain the correct (LO) finite terms for the process at hand. 
Large renormalisation-scale uncertainties could only be ameliorated by including more NLO matrix elements, and/or by improving or matching the underlying shower formalism to higher-logarithmic accuracy. Several approaches for the former are now emerging (and are available in PYTHIA, the most advanced being UNLOPS~\cite{Lonnblad:2012ix}), while the latter remains a  long-standing and highly non-trivial problem.

Our approach is based on a proposal first made in Ref.~\cite{Giele:2011cb},  which we have here proved to be valid to all orders in perturbation theory. We have also included several validations illustrating that the automated weight variations produced by our implementation do indeed reproduce the results of independent runs with the corresponding parameter changes. The formalism shares qualitative features with the proposal for ''boosting'' splitting probabilities in~\cite{Lonnblad:2012hz} and with the proposal for biasing photon emissions made in Ref.~\cite{Hoeche:2009xc}, and indeed for those purposes our approach reduces to those of~\cite{Hoeche:2009xc,Lonnblad:2012hz}.

Recently, two techniques for fast uncertainty variations for NLO calculations were presented~\cite{Bothmann:2015woa} in the context of the SHERPA event generator~\cite{Gleisberg:2008ta}, one based on interpolation grids and another based on analytically calculable weights. Our approach differs in several respects from both of these. Most importantly, our formalism applies to all orders rather than just to the first order of corrections, hence variations are performed all throughout the shower. Secondly, as we have shown, our strategy is formally exact (in the limit of infinitely many generated events), while the weights computed in \cite{Bothmann:2015woa} are only approximate in the shower context. An elaborate study of parton-shower uncertainties was also recently performed in the HERWIG context, using conventional methods (independent runs)~\cite{Bellm:2016rhh}\footnote{An older proposal was that of \cite{Stephens:2007ah}, which however was complicated by the fact that it did not rely directly on the veto algorithm to compute the variation weights.}.

We end by remarking on possible pathologies that can arise, and how best to deal with them.  If an event is very rare in the baseline sample but much more likely in a variation, the result will necessarily be a very large weight for that variation. Especially after cuts the statistical precision of the weighted samples can therefore be much lower than for the nominal ones. To address this, we recommend biasing the nominal sample to make the relevant rare occurrences more frequent. This has the additional benefit of improving the statistical precision also of the nominal weights in the tails of the distributions. Note however that the technology for combining the uncertainty variations with biases has not yet been implemented at the time of writing; we eagerly await feedback from the community on issues encountered in practical studies, on which to base the development of future capabilities and recommendations. 

Details of how to switch on the new automated framework in PYTHIA and how to define the list of uncertainty variations to be performed in an actual run have been included in a new HTML documentation file in the online set of PYTHIA documentation files. These technical specifications may change as the code evolves. The current version of these descriptions can be consulted at:
\url{http://home.thep.lu.se/~torbjorn/pythia82html/Welcome.html}, under the index heading ``Automated Shower Variations". 

\paragraph{Note added in proof:} during the completion of this work, we became aware of two complementary projects which allow to perform renormalisation-scale and PDF variations in a manner analogous to ours, implemented in the HERWIG and SHERPA generators respectively; see \cite{herwigPaper,Badger:2016bpw} for details. 

\paragraph{Acknowledgements:} The motivation to undertake this work arose from discussions at the Physics at TeV Colliders Workshop, Les Houches, June 2015. PS is supported in part by the Australian Research Council, contract FT130100744.  Fermilab is operated by the Fermi Research Alliance, LLC under Contract No. De-AC02-07CH11359 with the United States Department of Energy.

\bibliographystyle{utphys}
\bibliography{main}

\end{document}